\documentclass[11pt]{article}
\usepackage{amssymb}
\usepackage{graphicx}
\usepackage{multicol}
\usepackage{floatflt}

\advance\textheight by \topskip
\large\normalsize
\newcommand{\be}{\begin{equation}}
\newcommand{\ee}{\end{equation}}
\newcommand{\ba}{\begin{array}}
\newcommand{\ea}{\end{array}}

\newcommand{\eV}{{\rm eV}}

\def\Red  {}

\def\Black{}
\def\Blue {}

\newcommand{\MeV}{\,{\rm MeV}}

\newcommand{\PL}{Phys. Lett.}
\newcommand{\PR}{Phys. Rev.}

\def\circa#1{\,\raise.3ex\hbox{$#1$\kern-.75em\lower1ex\hbox{$\sim$}}\,}
\makeatletter
\def\art{\@ifnextchar[{\eart}{\oart}}
\def\eart[#1]#2#3#4#5#6{{\rm #2}, {\em #3 \bf #4} {\rm (#6) #5} ({\em #1})}
\def\hepart[#1]#2{{\rm #2, \em#1}}
\newcommand{\oart}[5]{{\rm #1}, {\em #2 \bf #3} {\rm (#5) #4}}

\newcounter{alphaequation}[equation]
\def\thealphaequation{\theequation\hbox to
0.6em{\hfil\alph{alphaequation}\hfil}}
\def\eqnsystem#1{
\def\@eqnnum{{\rm (\thealphaequation)}}
\def\@@eqncr{\let\@tempa\relax \ifcase\@eqcnt \def\@tempa{& & &} \or
  \def\@tempa{& &}\or \def\@tempa{&}\fi\@tempa
  \if@eqnsw\@eqnnum\refstepcounter{alphaequation}\fi
\global\@eqnswtrue\global\@eqcnt=0\cr}
\refstepcounter{equation} \let\@currentlabel\theequation \def\@tempb{#1}
\ifx\@tempb\empty\else\label{#1}\fi
\refstepcounter{alphaequation}
\let\@currentlabel\thealphaequation
\global\@eqnswtrue\global\@eqcnt=0 \tabskip\@centering\let\\=\@eqncr
$$\halign to \displaywidth\bgroup \@eqnsel\hskip\@centering
$\displaystyle\tabskip\z@{##}$&\global\@eqcnt\@ne
\hskip2\arraycolsep\hfil${##}$\hfil& \global\@eqcnt\tw@\hskip2\arraycolsep
$\displaystyle\tabskip\z@{##}$\hfil
\tabskip\@centering&\llap{##}\tabskip\z@\cr}
\def\endeqnsystem{\@@eqncr\egroup$$\global\@ignoretrue} \makeatother

\begin{document}

\centerline{hep-ph/0205261 \hfill CERN--TH/2002--109\hfill IFUP--TH/2002--22}
\vspace{5mm}
\Black
\vspace{0.5cm}
\centerline{\LARGE\bf\Red Which solar neutrino data favour the LMA solution?}

\medskip\bigskip\Black

\centerline{\large\bf 
A.\ Strumia$^{a,b}$, 
C.\ Cattadori$^{c,d}$, 
N.\ Ferrari$^c$, 
F.\ Vissani$^c$}\vspace{0.4cm}

\centerline{\em ${}^a$Dipartimento di Fisica dell'Universit\`a 
di Pisa and INFN}
 \vspace{0.1cm}
\centerline{\em ${}^b$Theoretical Physics Division, 
CERN, CH-1211 Gen\`eve 23, Suisse}
 \vspace{0.1cm}
\centerline{\em ${}^c$INFN, Laboratori Nazionali del 
Gran Sasso, I-67010 Assergi (AQ), Italy}
 \vspace{0.1cm}
\centerline{\em ${}^d$INFN Sezione di Milano, Via Celoria 
16, I-20133 Milano, Italy}
 \vspace{8mm}

\Blue\centerline{\large\bf Abstract}
\begin{quote}\large\indent
Assuming neutrino oscillations, global analyses of solar data
find that the LOW solution is significantly disfavoured,
leaving LMA as the best solution.
But the preference for LMA rests on three weak hints:
the spectrum of earth matter effects (Super-Kamiokande 
sees an overall day/night asymmetry only at $1\sigma$),
the Cl rate (but LMA and LOW predictions are 
both above the measured value),
the Ga rate (newer data decrease towards the LOW 
predictions both in GNO and SAGE).
Only new data will tell us if LMA is the true solution.

\Black
\end{quote}

\vspace{1cm}

\noindent
The recent SNO results~\cite{SNO} give a $5\sigma$ 
evidence for appearance of $\nu_{\mu,\tau}$
and increase the confidence in the standard solar model
prediction for the Boron flux.
There is also an indirect impact of these results 
on the determination of the true oscillation solution,
when they are combined with the other solar data~\cite{sunexp};
this suggests that the LMA solution is significantly favoured.
The question at stake is whether present data already
ensure the success of the research program of KamLAND
(that can see a clean signal in the LMA region),
or if we will measure the oscillation parameters only after
BOREXINO (that can see clean signals in the LOW or (Q)VO regions).

The global analyses of solar data agree 
that the LMA solution is significantly favoured.
It is clear why small-mixing-angle and purely-sterile oscillations
are `extremely disfavoured' or `excluded'.
E.g., SMA predicts $P_{ee}(E_\nu) = e^{-{\bar E}/E_\nu}$ 
where $\bar E$ is a constant at
energies probed by Super-Kamiokande (SK), and consequently 
a distortion of the 
energy spectrum, that is contradicted by the observations.
If SMA or sterile were the true solution, some well defined 
experimental result has to be wrong.

It is less clear which data significantly disfavour LOW.
Indeed, no data would be contradicted if LOW were the true 
solution, as we now discuss.
The SK and SNO rates alone cannot discriminate the two solutions:
in both cases it is possible to obtain 
$P_{ee}\approx 0.35$ at $E_\nu\sim 10\MeV$.
After ensuring this value, the various global analyses
(that use slightly different sets of data 
and estimation of errors, and do not include the 
recent GNO rate \cite{gno02})
obtain the best-fit difference
$$\Delta\chi^2 \equiv  \chi^2_{\rm LOW} - \chi^2_{\rm LMA} =
10.7~\cite{SNO},~4.7~\cite{Barger}
\hbox{(V1)},~8.1~\cite{update},
7.5~\cite{Band},~8.8~\cite{Bahcall},~9.2~\cite{Barger}\hbox{(V3)}
,~9.0~\cite{SK02},~13.3~\cite{SMI}
$$ 
as the result of a combination 
of 3 different pieces of data:
(1) the energy and zenith-angle spectrum at SK and SNO,
(2) the Chlorine rate, and (3) the Gallium rate.
Each of them only weakly prefers LMA to LOW, and 
when we combine all data together, 
a strong preference for LMA emerges. 
It is not possible to fully
isolate the relative importance
of the 3 data, since the best-fit points  
result from an interplay of all of them, in which 
a compromise is reached between obtaining a good fit of the rates, 
and avoiding too large spectral and day/night effects.
However, when we remove from
our fit\footnote{The present fit is performed as in~\cite{update}.
It takes into account the Chlorine and Gallium rates,
the full zenith angle and energy spectrum from SK
(using instead the older day/night spectra LOW would be less disfavoured)
and the NC and CC rate from SNO 
(not neglecting anti-correlations and spectral distortions).
We do not include the SNO energy spectrum (an approximated analysis shows
that it does not significantly affect any of our 
result of the fit in the
LMA, LOW, SMA, (Q)VO regions).
Up to updates and minor improvements,
errors on cross-sections and solar model predictions
(we use the BP00 model~\cite{BP00}) are computed following~\cite{LisiChiq}.
Survival probabilities in the sun are computed using the 
accurate approximations from~\cite{MSW}.
Earth matter effects are computed in the 
mantle/core approximation, improved by
using the average density appropriate for each 
trajectory as predicted in~\cite{PREM}.}
one of the 3 data previously
listed, the value $\Delta\chi^2 = 6.9$ that
we get diminishes to $4.6$, $5.9$, and $3$.

\begin{figure}
$$\includegraphics[width=16cm]{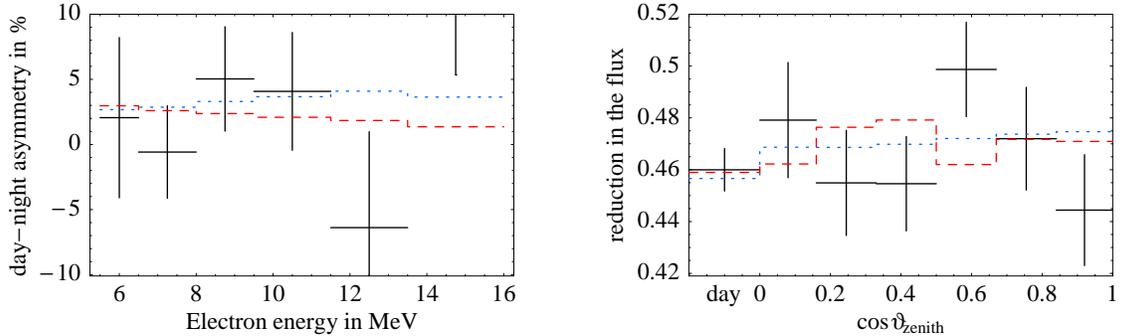}$$
\caption[]{\em SK energy and zenith-angle spectra (crosses), compared
with two representative LMA (blue dotted) 
and LOW (red dashed) oscillations.
LMA is slightly favoured.
 \label{fig:zenith}}
\end{figure}

\subsubsection*{The energy and zenith-angle spectrum}
SK and SNO data do not show a statistically 
significant hint for earth matter 
effects or for spectral distortions of Boron neutrinos.
The overall day/night asymmetries are 
$2.1 \pm 2.5\%$ at SK and $7.0\pm 5.1\%$ in CC events at SNO.
Both LMA and LOW can fit the overall 
day/night asymmetry, but they predict  
a different energy and zenith-angle
spectrum of matter effects, as 
illustrated in fig.~1\footnote{This figure 
has been obtained doing appropriate averages of the full
SK data; this can be done in a 
few sligtly different ways (our global fit 
employes the full data).
To optimize the visual appearance of fig 1b, we rescaled 
the value of the ${}^8$B flux for the 
LMA and LOW predictions.} 
for the points 
(very close to the LMA and LOW best-fit points) defined in table~1.
A look at SK and SNO data does not reveal any 
clear pattern; nevertheless, the global fit 
tells us that SK spectra favour the LMA point by 
$\Delta\chi^2_{\rm SK}\approx 3$.

\begin{table}[t]
$$\begin{array}{ccc|cc}
&\Delta m^2 & \tan^2\theta & \hbox{Ga rate in SNU} & 
\hbox{Cl rate in SNU}\\ \hline
\hbox{LMA} & 10^{-4.2}\eV^2 & 10^{-0.4} & 70.2& 2.85 \\
\hbox{LOW} & 10^{-7.1}\eV^2 & 10^{-0.2} & 64.7 & 3.17\\ \hline
\multicolumn{3}{c|}{\hbox{experimental value}} & 70.8\pm 4.4 & 2.56 \pm 0.23
\end{array}$$
\caption{\em Predicted central values of Ga and Cl low-energy neutrino rates.
We only show the experimental errors.
Smaller errors from solar models, earth models, 
cross sections, oscillation fits
are strongly correlated and cannot be presented in a simple and reliable way.}
\end{table}

\subsubsection*{The Chlorine rate}
Both LMA and LOW solutions predict an increase of 
the survival probability $P_{ee}$ at energies below the SK and SNO thresholds.
In the LMA case this happens because neutrinos of lower energy
do not experience the matter resonance in the 
sun, so that $P_{ee}$ increases from
$\sin^2\theta$ (adiabatic MSW effect) to 
$1-\frac{1}{2}\sin^2 2\theta$ (averaged vacuum oscillations).
In the LOW case the increase of $P_{ee}$ is due to 
a large $\nu_e$ regeneration induced by earth matter effects.
LMA and LOW predict a different increase of $P_{ee}$, and this can 
be tested by experiments sensitive to lower energy neutrinos.

The Cl experiment has some limited sensitivity to low energy 
neutrinos: roughly $80\%$ of the Cl rate is due to the Boron neutrinos,
that above $6\MeV$ are better studied by SK and SNO
(the Cl rate has not yet been calibrated, though theoretical 
cross sections seem to be sufficiently reliable).
After the recent SNO measurement of the Boron flux,
we can use the Cl rate to get better information on
low energy neutrinos. 
As exemplified in table~1,
the best-fit LOW prediction is larger than 
the LMA prediction (computed at the central value of 
the detection cross section,
and of solar and earth models),
and both are larger than the measured Cl rate.
This fact amplifies the impact in the $\chi^2$ 
beyond the sensitivity of the experiment\footnote{This 
can be illustrated by a simple example. 
A $1\sigma$ difference between two solutions 
with a $2.5\sigma$ pull and a $1.5\sigma$ pull
gives $\Delta\chi^2 = 2.5^2-1.5^2 = 2^2$, rather than $1^2$.
This is why, when pulls are `too' large, 
one can prefer to treat the data in
some non-standard more conservative way.},
that remains anyhow mild.

\renewcommand\arraystretch{1.2}
\begin{table}[t]
\label{galliumexp}
\begin{center}
\begin{tabular}{|c|c|c c|}
\hline 
SAGE & Jan 1990 --- Dec 2001 & Jan 1990 --- Oct 
1999 & Apr 1998 --- Dec 2001\\
\hline
N.\ runs & 92 & 70 & 35\\
$R_{\nu}$ & 70.8$^{+5.3}_{-5.2}\hbox{(stat)}^{+3.7}_{-3.2}\hbox{(syst)}$ &
75.4$^{+7.0}_{-6.8}\hbox{(stat)}^{+3.5}_{-3.0}\hbox{(syst)}$ & 
67$^{+7.0}_{-7.0}\hbox{(stat)}$\\
\hline\hline
GALLEX-GNO & May 1991 --- Dec 2001 & 
May 1991 --- Jan 1997 & Apr 1998 --- Dec 2001 \\
\hline
N.\ runs & 108 & 65 & 43\\
$R_{\nu}$ & 70.8$^{+4.5}_{-4.5}\hbox{(stat)}^{+3.8}_{-3.8}\hbox{(syst)}$
&77.5$^{+6.2}_{-6.2}\hbox{(stat)}^{+4.3}_{-4.7}\hbox{(syst)}$ &
65.2$^{+6.4}_{-6.4}\hbox{(stat)}^{+3.0}_{-3.0}\hbox{(syst)}$ \\
\hline
\end{tabular}
\caption{\em Solar neutrino interaction rates measured by SAGE, GALLEX, 
GNO Gallium experiments.}
\end{center}
\end{table}

\subsubsection*{The Gallium rate}
The Ga rate is dominantly due to low energy neutrinos;
the Boron contribution is about $10\%$.
As exemplified in table~1, the best-fit 
LOW prediction is somewhat lower than the LMA prediction,
that is closer to the measured value.
The preference for LMA versus LOW decreases down to
$\Delta\chi^2\approx 3$ if the Ga rate is not taken 
into account in a global fit. 

Table~2 shows the solar neutrino interaction rates
measured by SAGE and GALLEX-GNO Gallium experiments since 1991.
The most recent values of the
solar neutrino interaction rates 
are lower than the earlier ones, 
both in SAGE~\cite{sage99,sage00,sage02} 
and in GNO \cite{gallex99,gno1,gno01,gno02}:
$$
\mbox{[Ga $1^{\rm st}$ period]}=76.4\pm 5.4\mbox{ SNU},\ \ \
\mbox{[Ga $2^{\rm nd}$ period]}=66.1\pm 5.3\mbox{ SNU}.
$$
Of course, these two values are experimentally and statistically
consistent, as the rate measured in the second period
(April 1998 -- December 2001) is just 1.5 standard 
deviations below the rate measured in the
former period (1991--1997).
For the sake of discussion, it has to be recalled  
that SAGE reported a problem in
data acquisition (DAQ) program \cite{sage02}, that 
strongly affected the years 1996--1999. To solve the problem, 
an ``a posteriori'' correction was applied~\cite{sage02}.
On the other side, GNO completely renewed electronic 
and DAQ \cite{gno1}, and succeeded to
reduce the already low background of about $30\%$~\cite{gno01}
in the analysis region---mainly, in the low energy $L$ window.
Finally both experiments in last period refined the 
systematics; as a results, the rate 
has larger statistical but lower systematic error
in the second period. To conclude, it is a fact that both 
Gallium  experiments improved with time.

\begin{figure}[t]
$$\includegraphics[width=8cm]{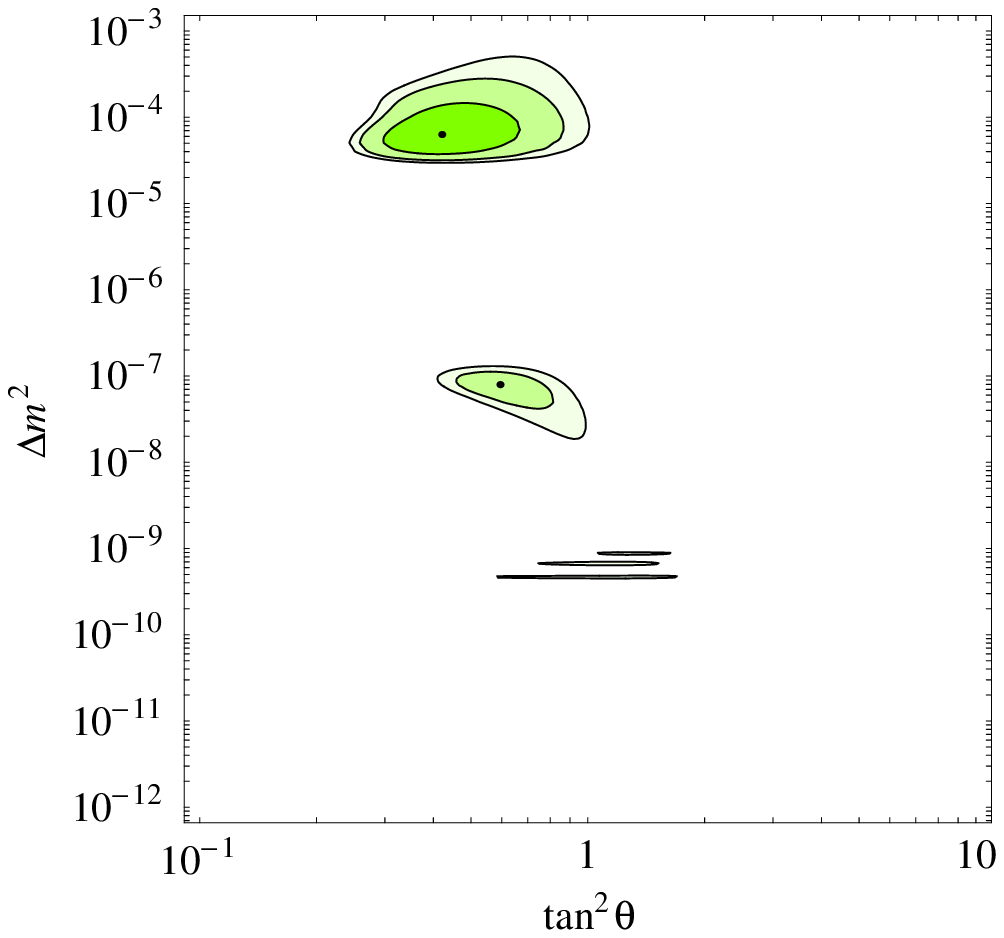}
\includegraphics[width=8cm]{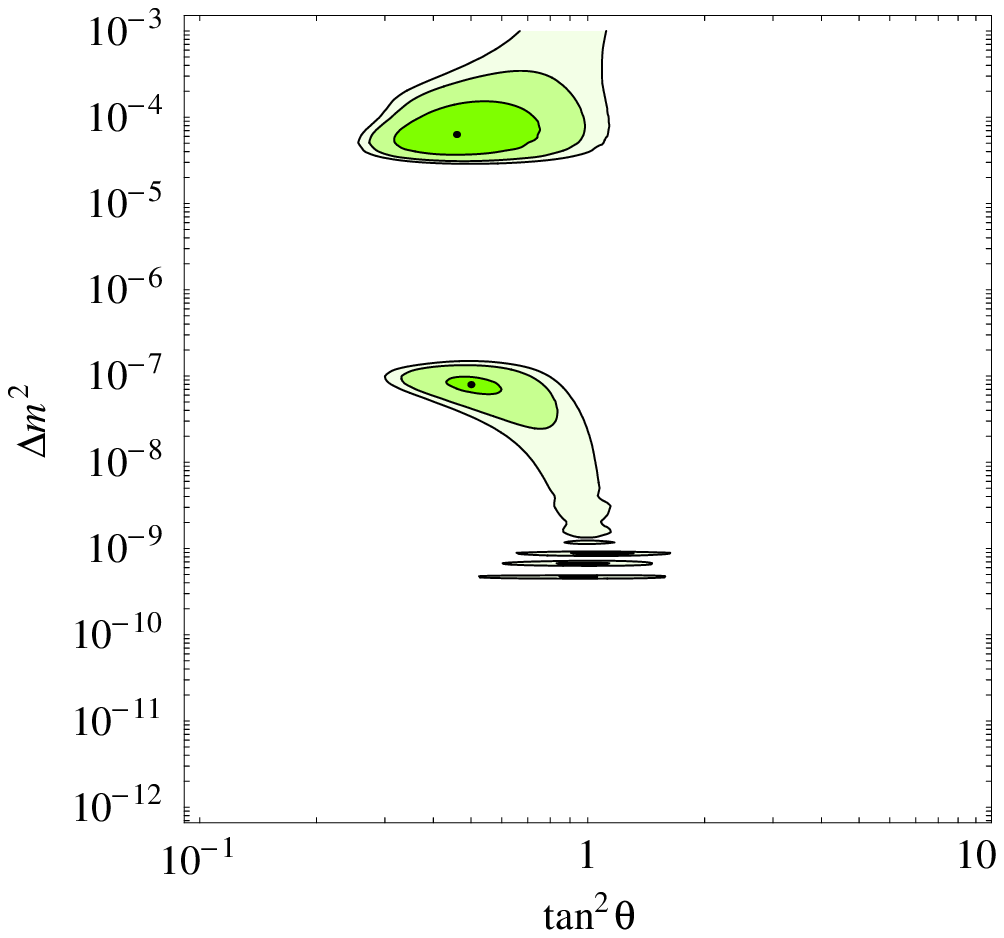}$$
\vspace{-10mm}
\caption[]{\em Global fit of solar data at $90,99,99.73\%$ {\rm CL}
using  all Gallium data ($70.8\pm 4.4$ SNU), fig.~a,
and only the newer Gallium data ($66.1\pm 5.3$ SNU), fig.~b.
\label{fig:fits}}
\end{figure}

If the difference between the
rates measured in the two periods is due
to a statistical fluctuation, or 
if it cannot be proved  that other 
systematic effects are present, the right  
thing to do is to average the two values as done 
in all global analyses~\cite{Barger,update,Band,Bahcall,SK02,SMI}.
The result of such a kind of 
global fit significantly disfavours the 
LOW solution, as shown in fig.~\ref{fig:fits}a, though
the inclusion of the GNO data reduces 
slightly the difference: we find $\Delta\chi^2 \approx 6.9$.

On the other side, the result of a fit 
that uses {\em only} data from the second period
is the one shown in fig.~\ref{fig:fits}b:
the LOW solution turns out to be less 
disfavoured, $\Delta\chi^2 \approx 3.9$.

\paragraph{Conclusions}
We have performed a global fit of solar data, including the
recent GNO result.
We have shown that the 
strong preference for LMA 
rests on three weak hints.
Only new data will 
tell us  if the solution is LMA or LOW/(Q)VO; 
the second eventuality would be not surprising,
as it would not contradict any present data.

\paragraph{Acknowledgements} We thank Carlos Pe\~na-Garay for many 
useful discussions.

\footnotesize
\begin{multicols}{2}

\end{multicols}
\end{document}